\documentclass[aps,pra,twocolumn,showpacs,superscriptaddress,longbibliography]{revtex4-2}
\usepackage{graphicx} 
\usepackage{amsmath}
\usepackage{graphicx,epstopdf}
\usepackage{gensymb}
\newcommand{\be}{\begin{equation}}
\newcommand{\ee}{\end{equation}}
\newcommand{\bea}{\begin{eqnarray}}
\newcommand{\eea}{\end{eqnarray}}
\newcommand{\bse}{\begin{subequations}}
	\newcommand{\ese}{\end{subequations}}

\usepackage{color}
\usepackage[colorlinks,bookmarks=false,citecolor=darkblue,linkcolor=red,urlcolor=blue]{hyperref}
\def\cred{\color{red}}
\definecolor{darkred}{rgb}{0.7,0.0,0.0}

\definecolor{darkblue}{rgb}{0,0.02,0.45}

\definecolor{darkgreen}{rgb}{0.02,0.45,0.0}

\definecolor{violet}{rgb}{0.8,0.2,0.6}

\begin{document}

\title{Low-dimensional magnetism of BaCuTe$_{2}$O$_{6}$}

\author{P. Bag}
\author{N. Ahmed}
\author{Vikram Singh}
\affiliation{School of Physics, Indian Institute of Science
Education and Research Thiruvananthapuram-695551, India}
\author{M. Sahoo}
\affiliation{Department of Physics, University of Kerala, Kariavattom, Thiruvananthapuram-695581, India}
\author{A. A. Tsirlin}
\email{altsirlin@gmail.com}
\affiliation{Experimental Physics VI, Center for Electronic Correlations and Magnetism, Institute of Physics, University of Augsburg, 86135 Augsburg, Germany}
\author{R. Nath}
\email{rnath@iisertvm.ac.in}
\affiliation{School of Physics, Indian Institute of Science
	Education and Research Thiruvananthapuram-695551, India}
\date{\today}

\begin{abstract}
X-ray diffraction, thermodynamic measurements, and density-functional band-structure calculations are used to study the magnetic behavior of BaCuTe$_{2}$O$_{6}$, a member of the $A$CuTe$_2$O$_6$ structural family that hosts complex three-dimensional frustrated spin networks with possible spin-liquid physics. Temperature-dependent magnetic susceptibility and heat capacity of the Ba compound are well described by the one-dimensional spin-$\frac12$ Heisenberg chain model reminiscent of the Sr analog SrCuTe$_2$O$_6$. While the intrachain coupling $J/k_{\rm B}\simeq 37$\,K is reduced compared to 49\,K in the Sr compound, the N\'eel temperature increases from 5.5\,K (Sr) to 6.1\,K (Ba). Unlike the Sr compound, BaCuTe$_2$O$_6$ undergoes only one magnetic transition as a function of temperature and shows signatures of weak spin canting. We elucidate the microscopic difference between the Sr and Ba compounds and suggest that one of the interchain couplings changes sign as a result of negative pressure caused by the Sr/Ba substitution. The N\'eel temperature of BaCuTe$_2$O$_6$ is remarkably insensitive to the magnetic dilution with Zn$^{2+}$ up to the highest reachable level of about 20\,\%.
\end{abstract}

\maketitle

\section{Introduction}
Frustrated magnetic networks host unusual ground states and non-trivial dynamical phenomena that become intricate especially in the case of quantum spins~\cite{Savary2017,Knolle2019,Broholm2020}. Compounds based on the spin-$\frac12$ Cu$^{2+}$ ions offer one of the best playgrounds for the experimental realization of such networks. Different types of crystal structures give access to multiple interaction geometries, including kagome lattices of interacting spins~\cite{Mendels2016,Norman2016}. Layered kagome compounds can even be found in nature as copper minerals~\cite{Inosov2018}, but their three-dimensional analogs -- material realizations of the hyperkagome spin lattice -- remain scarce. The only example in the cuprate family is PbCuTe$_2$O$_6$~\cite{Koteswararao2014,Khuntia2016}, where signatures of a three-dimensional quantum spin liquid with fractional spinon excitations were recently reported~\cite{Chillal2020a}.

PbCuTe$_2$O$_6$ has a fairly complex cubic crystal structure. The hyperkagome lattice formed by second-neighbor interactions is augmented by interactions between first and third neighbors~\cite{Chillal2020a}. When Sr substitutes for Pb, the crystal structure and even the cubic lattice parameter remain nearly unchanged, but the magnetic behavior is modified drastically. SrCuTe$_2$O$_6$ is a spin-chain compound with a robust magnetic order at low temperatures~\cite{Ahmed2015,Saeaun2020,Chillal2020b}. Predominant third-neighbor interactions give rise to spin chains along each of the three main cubic directions~\cite{Ahmed2015}. The presence of multiple phase transitions~\cite{Ahmed2015,Koteswararao2015}, complex temperature-field phase diagram~\cite{Chillal2020b}, and non-collinear magnetic order~\cite{Saeaun2020,Chillal2020b} witness frustrated interchain interactions that may be residues of the strong frustration present in PbCuTe$_2$O$_6$. 

The drastically different behavior of the Pb and Sr compounds suggests an extreme sensitivity of this structure type to chemical substitutions. This fact motivated us to prepare and explore BaCuTe$_2$O$_6$, another member of the same family, {\cred and attempt its further tuning by Zn doping that may introduce both chemical pressure and magnetic dilution}.

BaCuTe$_2$O$_6$ retains the cubic crystal symmetry and overall structural geometry of the $A$CuTe$_2$O$_6$ compounds, but shows a notable lattice expansion, with the room-temperature cubic lattice parameter increasing from $a=12.473$\,\r A~\cite{Wulff1997} (Sr) and $a=12.49$\,\r A~\cite{Koteswararao2014} (Pb) to $a=12.87$\,\r A (Ba). We demonstrate that, despite this effective negative pressure, BaCuTe$_2$O$_6$ strongly resembles SrCuTe$_2$O$_6$ and hosts weakly coupled spin chains due to predominant magnetic interactions between third neighbors. Interestingly, substituting Ba for Sr decreases the intrachain coupling but increases the N\'eel temperature. We combine this observation with an \textit{ab initio} microscopic analysis to identify relevant magnetic interactions and their changes upon chemical substitutions. {\cred We also show a remarkable insensitivity of the N\'eel temperature to magnetic dilution as another experimental witness of the underlying frustration in this family of compounds.}

\section{Methods}
\label{sec:methods}
Polycrystalline samples of Ba(Cu$_{1-x}$Zn$_{x}$)Te$_{2}$O$_{6}$ ($x$ = 0~\textendash~0.20) were synthesized following the same procedure adopted for the synthesis of SrCuTe$_{2}$O$_{6}$~\cite{Ahmed2015}. The initial reactants BaCO$_{3}$ (Aldrich, 99.995\%), CuO (Aldrich, 99.999\%), TeO$_{2}$ (Aldrich, 99.995\%), H$_{2}$TeO$_{4}$.2H$_{2}$O (Alfa Aesar, 99\%), and ZnO (Aldrich, 99.99\%) were taken in stoichiometric ratios required for the synthesis of BaCuTe$_{2}$O$_{7}$ and its Zn doped samples. For $x = 0$, the reactants were ground thoroughly, pelletized, and fired at 670~$\degree$C for five days in flowing argon atmosphere with several intermediate grindings. For the doped samples, the firing temperature was varied from 670~$\degree$C to 720~$\degree$C.

Similar to SrCuTe$_{2}$O$_{6}$, the flowing argon provides a reducing atmosphere which leads to the loss of oxygen and hence the formation of BaCuTe$_{2}$O$_{6}$ instead of BaCuTe$_{2}$O$_{7}$. The powder x-ray diffraction (XRD) pattern was recorded at room temperature using a PANalytical powder diffractometer (Cu\textit{K}$_{\alpha}$ radiation, $\lambda_{\rm avg}\simeq 1.5418$~{\AA}). {\cred The samples were found to be single-phase up to $x = 0.18$, while for $x = 0.20$ BaCuTe$_{2}$O$_{7}$ emerged as a minor impurity phase.} The amount of the impurity phase for $x = 0.20$ was found to be very small (about $\sim 2$\%). Our repeated attempts to achieve higher doping levels by increasing or lowering the firing temperature were unsuccessful. 

\begin{figure}
\includegraphics{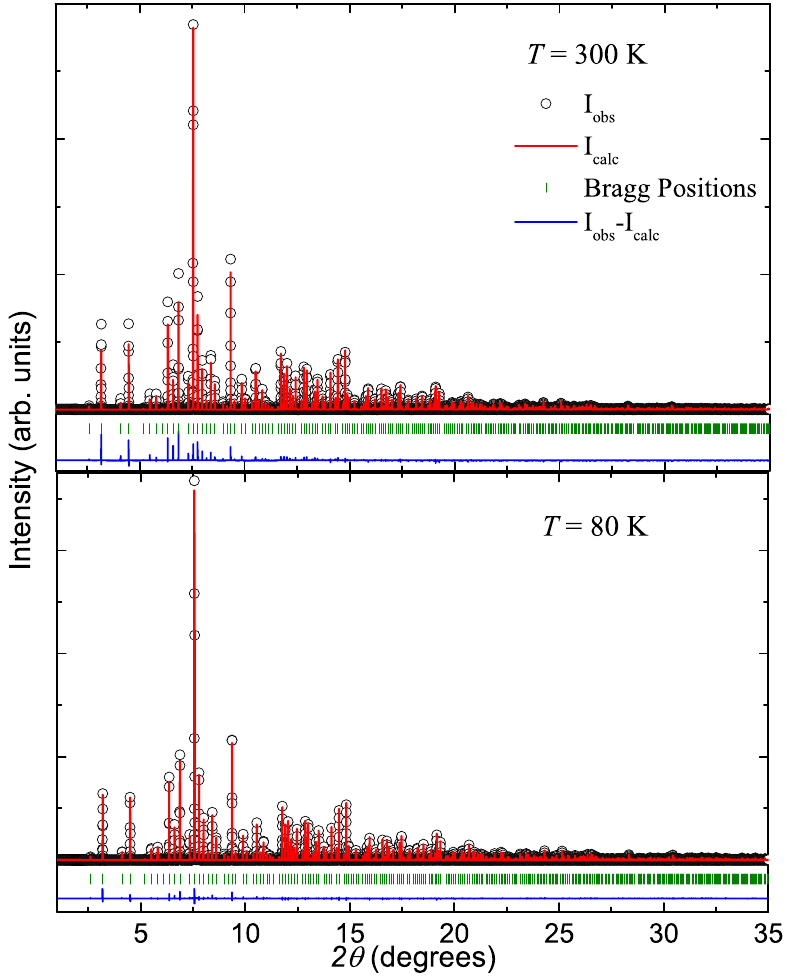}
\caption{\label{Fig1}
Synchrotron x-ray diffraction data of BaCuTe$_2$O$_6$ at 300~K and 80~K. The solid lines denote the Rietveld fits of the data. The Bragg peak positions are indicated by green vertical bars, and the bottom blue line shows the difference between the experimental and calculated intensities. {\cred The refinement residuals $R_I/R_p$ are 0.028/0.092 at 300\,K and 0.016/0.076 at 80\,K.}}
\end{figure}

\begin{figure}
	\includegraphics[width=8.8cm]{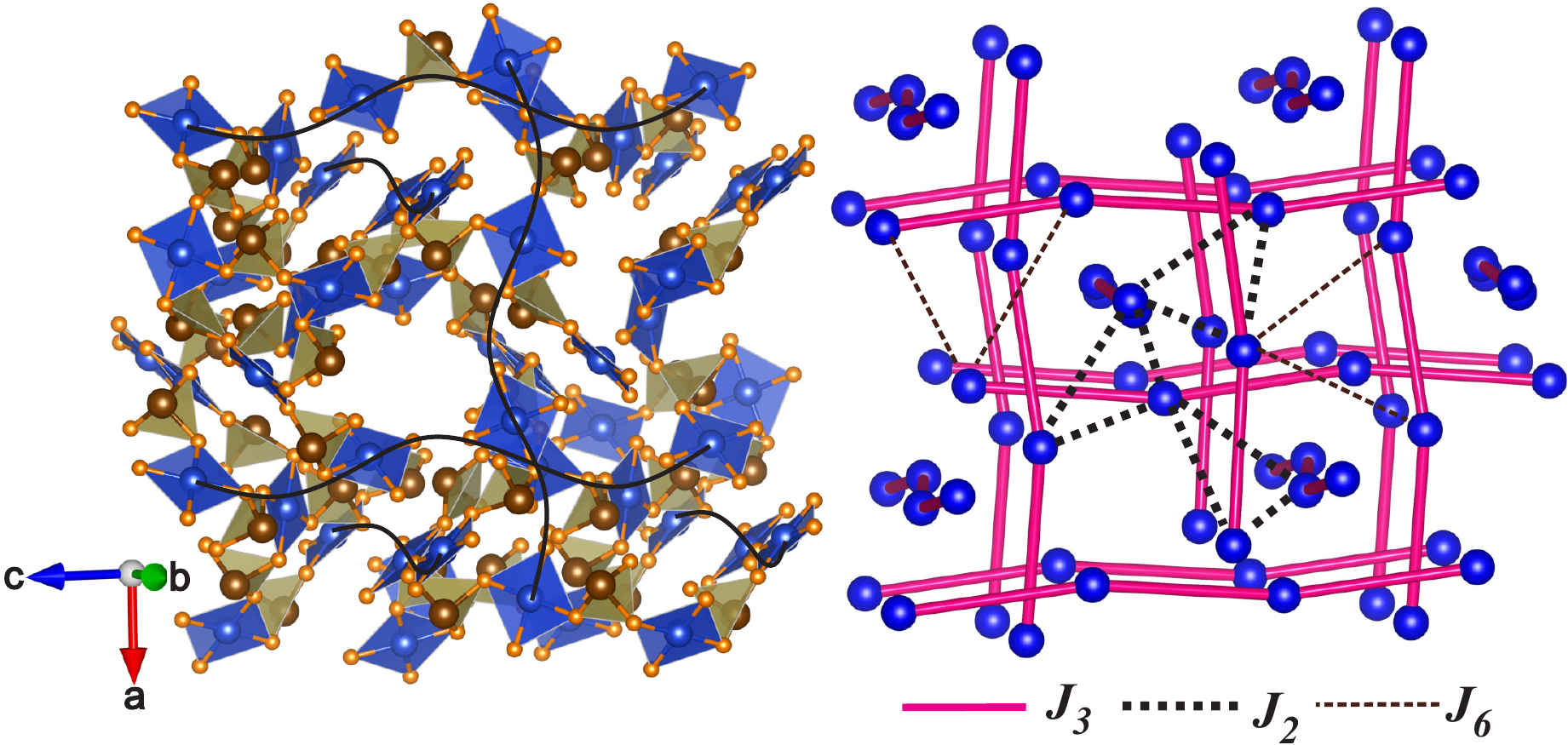}
	\caption{\label{Fig2}
		Left panel: Crystal structure of BaCuTe$_2$O$_6$, with black curved lines denoting the spin chains. The blue square planes are the CuO$_{4}$ plaquettes, while the TeO$_{3}$ units are shown in green color. The Ba atoms are omitted for clarity. Right panel: spin lattice of BaCuTe$_2$O$_6$ composed of crossed spin chains (solid red lines) of Cu$^{2+}$ ions. The dominant interchain couplings $J_2$ and $J_6$ forming frustrated loops are also shown. \texttt{VESTA} software was used for crystal structure visualization~\cite{vesta}.}
\end{figure}

For structure determination, high-resolution synchrotron XRD data were collected at the ID22 beamline of ESRF using the wavelength of 0.4107\,\r A. The powder sample placed into a thin-wall borosilicate capillary was cooled down with liquid-nytrogen cryostream and spun during the measurement. Diffracted signal was measured by 9 scintillation detectors, each preceded by a Si (111) analyzer crystal. Rietveld refinements of the XRD data were performed using \texttt{JANA2006} and \texttt{FULLPROF} software packages~\cite{Carvajal1993,Petvrivcek345}.

Magnetic susceptibility $\chi$ ($\equiv M/H$, $M$ is the magnetization and $H$ is the applied field) was measured as a function of temperature (2~K~$\leq T \leq 350$~K) and magnetic field (0~T~$\leq H \leq 9$~T) using a vibrating sample magnetometer (VSM) attachment to the Physical Property Measurement System (PPMS, Quantum Design). Heat capacity $C_{\rm p}$ was measured on a small piece of the sintered pellet via the relaxation technique using the heat capacity option of PPMS.


Exchange parameters were calculated for the spin Hamiltonian
\begin{equation}
 \mathcal H=\sum_{\langle ij\rangle} J_{ij}\mathbf S_i\mathbf S_j,
\label{eq:ham}\end{equation}
where the summation is over lattice bonds $\langle ij\rangle$. Exchange integrals $J_{ij}$ were extracted via density-functional (DFT) band structure calculations performed using the \texttt{FPLO} code~\cite{fplo} within generalized gradient approximation (GGA)~\cite{pbe96} for the exchange-correlation potential. The mean-field GGA+$U$ approach was used to treat correlation effects in the Cu $3d$ shell, with the Hubbard repulsion parameter $U_d=8.5$\,eV, Hund's coupling $J_d=1$\,eV, and double-counting correction in the atomic limit~\cite{Mazurenko2014,Nath2014a}. All calculations were performed for the crystallographic unit cell with 120 atoms, and reciprocal space was sampled by 64 \textbf{k}-points in the symmetry-irreducible part of the first Brillouin zone. Experimental structural parameters were used, as further described in Sec.~\ref{sec:dft}.

\section{Results}
\subsection{Crystal Structure}
To the best of our knowledge, synthesis and characterization of BaCuTe$_2$O$_6$ have not been reported previously. To establish the crystal structure of this compound, Rietveld refinement of the synchrotron XRD data was performed using the structural data of SrCuTe$_{2}$O$_{6}$ as the starting model~\cite{Wulff1997}. Figure~\ref{Fig1} presents the synchrotron XRD data at two different temperatures along with the Rietveld fits. The results are listed in Table~\ref{tab:coordinates} and suggest close similarity to the Sr compound. At 80\,K, all atomic displacement parameters are reduced with respect to their room-temperature values, indicating the absence of structural disorder. The incorporation of Ba has the effect of negative pressure and increases the Cu--Cu distances compared to SrCuTe$_2$O$_6$ (see also Table~\ref{tab:exchange}).

\begin{table}
\setlength{\tabcolsep}{0.1cm}
\caption{\label{tab:coordinates}
Atomic positions and isotropic atomic displacement parameters ($U_{\rm iso}$, in units of $10^{-2}$~\AA$^{2}$) for BaCuTe$_2$O$_6$ at 80~K (upper lines) and 300~K (lower lines) obtained from the refinement of the synchrotron data. The cubic lattice parameter is $a=12.8296(1)$\,\r A at 80\,K and $a=12.8699(1)$\,\r A at 300\,K. The space group is $P4_132$. All the sites are fully occupied.}
\label{Positions}
\begin{tabular}{ccccccc}
		\hline \hline 
		& Atom      & Wyckoff     & $x/a$     & $y/a$    & $z/a$      & $U_{\rm iso}$ \\
		& &position & & & & \\\hline
		&Ba1  &$8c$ &0.05757(3) &0.05757(3)  &0.05757(3) &0.35(2) \\
		& & &0.05825(4) &0.05825(4) &0.05825(4) &1.29(2)\\[0.2cm]
		&Ba2 &$4b$ &0.375 &0.625 &0.125 &0.41(2)\\
		& & &0.375 &0.625 &0.125 &1.33(3)\\[0.2cm]
		&Cu &$12d$ &0.47429(7) &0.875 &0.27571(7) &0.37(3)\\
		& & &0.47381(9) &0.875 &0.27619(9) &1.17(2)\\[0.2cm]
		&Te &$24e$ &0.33709(4) &0.91707(4) &0.06354(4) &0.39(1)\\
		& & &0.33709(4) &0.91657(5) &0.06425(5) &1.31(1)\\[0.2cm]
		&O1 &$24e$ &0.3763(3) &0.8256(3) &0.1734(3) &0.29(1)\\
		& & &0.3760(4) &0.8254(4) &0.1733(4) &1.01(1)\\[0.2cm]
		&O2 &$24e$ &0.2664(3) &0.8115(3) &$-0.0121(3)$ &0.52(1)\\
		& & &0.2673(4) &0.8117(4) &$-0.0163(4)$ &1.82(1)\\[0.2cm]
		&O3 &$24e$ &0.2259(3) &0.9751(3) &0.1360(4) &0.89(1)\\
		& & &0.2265(4) &0.9746(4) &0.1376(4) &2.39(1)\\
		\hline
		\hline
\end{tabular}
\end{table}


Figure~\ref{Fig2} shows the crystal structure of BaCuTe$_2$O$_6$. The CuO$_4$ plaquettes are linked into a 3D network via TeO$_3$ trigonal pyramids. This geometrical arrangement allows for several interaction topologies, including isolated triangles formed by the nearest-neighbor interaction $J_1$, the hyperkagome lattice formed by the second-neighbor interaction $J_2$, and crossed spin chains formed by the third-neighbor interaction $J_3$ (see also Fig.~\ref{Fig10}). This latter interaction is predominant in BaCuTe$_2$O$_6$, as we show in the following.

\begin{figure}[h]
	\includegraphics[] {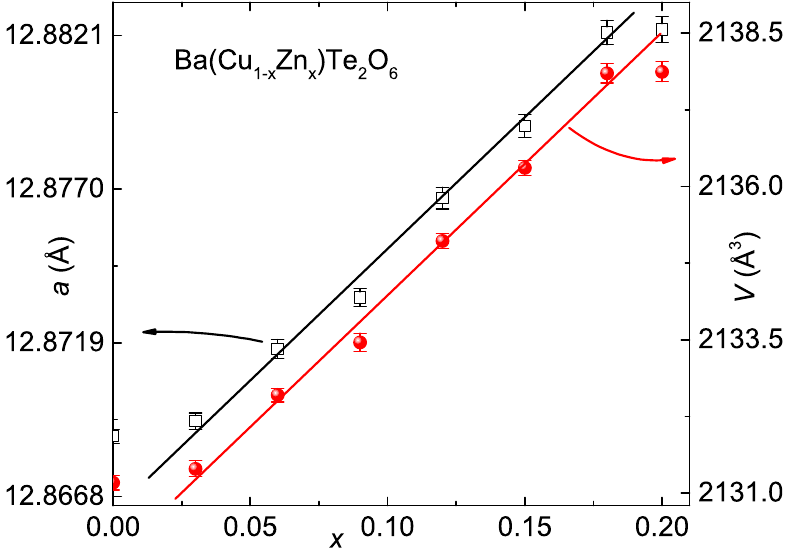}
	\caption{\label{Fig3} Variation of lattice constant ($a$) and unit cell volume ($V$) with Zn$^{2+}$ concentration ($x$), obtained from the Rietveld refinement of the powder XRD data at room temperature for Ba(Cu$_{1-x}$Zn$_{x}$)Te$_{2}$O$_{6}$. Solid lines are the linear fits as per the Vegard's law.}
\end{figure}
We have also prepared Zn-doped samples that preserve the cubic symmetry up to the highest doping level of $x\simeq 0.2$. The lattice constant ($a$) increases almost linearly but weakly with Zn concentration ($x$) (see Fig.~\ref{Fig3}) 
This leads to an overall linear increase in the unit cell volume ($V$). Though the change is marginal, the increasing trend resembles other Zn-doped Cu$^{2+}$ compounds~\cite{Ranjith2015} and complies with the slightly larger ionic radius of Zn$^{2+}$ (0.6~\AA) compared to Cu$^{2+}$ (0.57~\AA).


\subsection{Magnetic Susceptibility}
\begin{figure}
\includegraphics{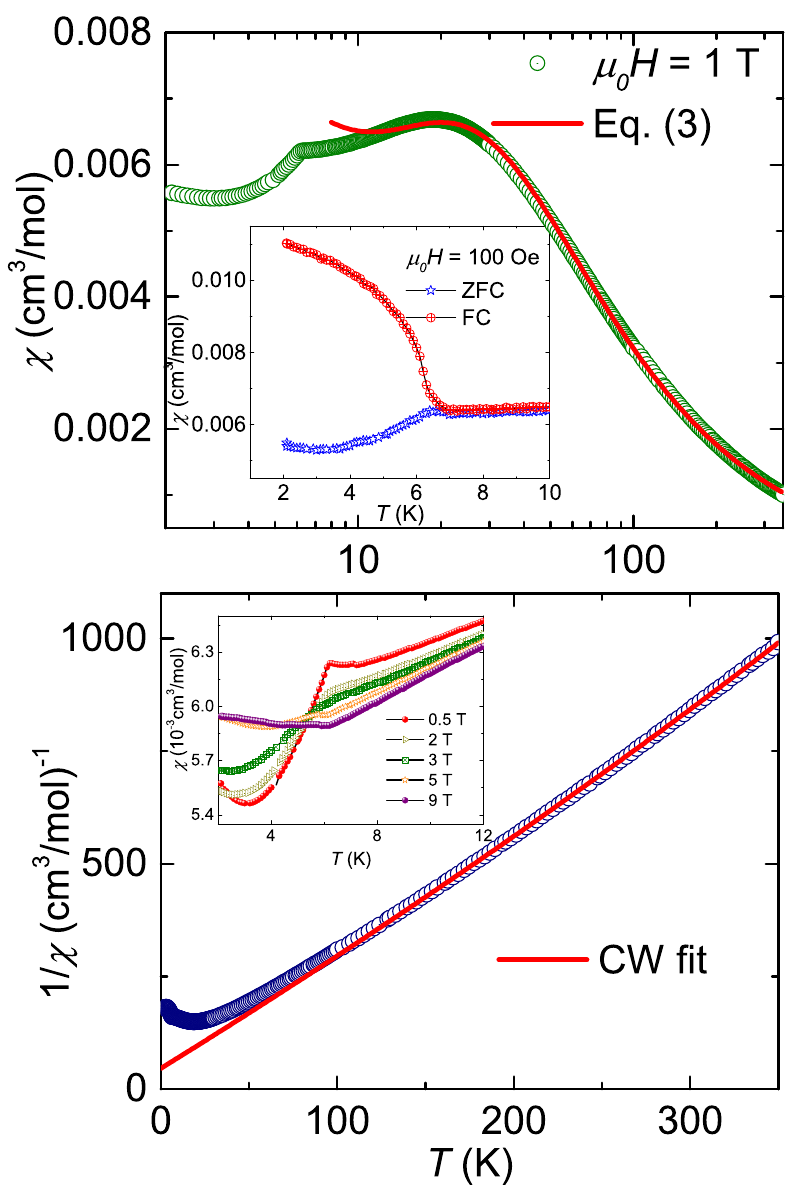}
\caption{\label{Fig4}
Upper panel: $\chi(T)$ of BaCuTe$_2$O$_6$ measured in an applied field of $\mu_0H=1$~T. The solid red line represents the 1D model fit [Eq.~\eqref{Johnston}] as described in the text. Inset: ZFC and FC susceptibilities measured at $\mu_0H = 0.01$~T in the low-temperature regime. Lower panel: Inverse susceptibility $1/\chi$ as a function of temperature. The solid line is the CW fit using Eq.~\eqref{cw}. Inset: Low-temperature $\chi(T)$ of BaCuTe$_{2}$O$_{6}$ measured at different applied fields up to 9~T.}
\end{figure}

Temperature-dependent magnetic susceptibility $\chi(T)$ measured in an applied magnetic field $\mu_0H=1$~T is shown in the upper panel of Fig.~\ref{Fig4}. It follows a Curie-Weiss (CW) behavior in the high-temperature regime. A broad maximum at $T_{\chi}^{\rm max} \simeq 20$~K is the signature of short-range order, typically observed in low-dimensional antiferromagnetic (AFM) spin systems. The temperature corresponding to the broad maximum is a measure of the strength of the dominant AFM exchange interaction~\cite{Johnston2000,Bonner1964,Eggert1994}. With decreasing temperature, a sharp peak reminiscent of a transition toward magnetic long-range-order (LRO) is observed at $T_{\rm N}~\simeq 6.1$~K. When the temperature is lowered further, a small Curie-like upturn in $\chi(T)$ is seen, which is due to the small amount of extrinsic paramagnetic impurities or intrinsic defects present in the sample.

The $\chi(T)$ data in the high-temperature region were fitted by the following expression:
\begin{equation}
\chi(T)=\chi_0+\frac{C}{T+\theta_{\rm CW}}.
\label{cw}
\end{equation}
Here $\chi_0$ is the temperature-independent susceptibility that includes contributions of core diamagnetism ($\chi_{\rm core}$) and Van-Vleck paramagnetism ($\chi_{\rm VV}$). The second term is the CW law with $C$ being the Curie constant and $\theta_{\rm CW}$ the characteristic Curie-Weiss temperature. Our fit above 200~K (lower panel of Fig.~\ref{Fig4}) yields $\chi_0 \simeq -1.19 \times 10^{-4}$~cm$^{3}$/mol, $C \simeq 0.41$~cm$^{3}$\,K/mol, and $\theta_{\rm CW} \simeq 18.97$~K. The value of $C$ corresponds to an effective moment $\mu_{\rm eff} \simeq 1.82~\mu_{\rm B}$, which is slightly higher than the spin-only value of $\mu_{\rm eff} = g\sqrt{S(S+1)}\mu_{\rm B} \simeq 1.73~\mu_{\rm B}$ for Cu$^{2+}$ ($S=\frac12$) and $g=2$. This is consistent with the $g$-value slightly above 2.0, as typical for Cu$^{2+}$ compounds~\cite{Nath2014a,Janson2011}. $\chi_{\rm core}$ of BaCuTe$_{2}$O$_{6}$ was calculated to be $-1.43 \times 10^{-4}$~cm$^{3}$/mol by adding the $\chi_{\rm core}$ of individual ions Ba$^{2+}$, Cu$^{2+}$, Te$^{4+}$, and O$^{2-}$~\cite{Selwood1956}. $\chi_{\rm VV}$ was calculated by subtracting $\chi_{\rm core}$ from $\chi_0$ and equals $\sim 2.3 \times 10^{-5}$~cm$^{3}$/mol. This value of $\chi_{\rm VV}$ is comparable to the other cuprate compounds like Sr$_2$CuO$_3$~\cite{Motoyama1996}, Sr$_2$Cu(PO$_4$)$_2$~\cite{Nath2005}, and PbCu$_{3}$TeO$_{7}$~\cite{Koteswararao2013}. The value of $\theta_{\rm CW}$ is positive and comparable in magnitude to $T_{\chi}^{\rm max}$, which implies that the dominant exchange interaction between the Cu$^{2+}$ ions is AFM in nature.

Magnetic susceptibility was further analyzed using the expression
\begin{equation}
 \chi(T)=\chi_0+\frac{C_{\rm imp}}{T} + \chi_{\rm spin}(T),
 \label{Johnston}
\end{equation}
where the second term is the Curie law that account for paramagnetic impurities potentially present in the sample, and $C_{\rm imp}$ gauges the impurity concentration. The third term $\chi_{\rm spin}(T)$ was taken as the expression for the spin susceptibility of a uniform 1D Heisenberg spin-$\frac{1}{2}$ AFM chain given by Johnston $\textit{et al}$~\cite{Johnston2000}. This expression is valid over a wide temperature range ($5 \times 10^{-25} \leq k_{\rm B}T/J \leq 5$). The fit above 10\,K returned $\chi_0 \simeq -3.76 \times 10^{-5}$~cm$^{3}$/mol, $C_{\rm imp} \simeq 0.013$~cm$^{3}$~K/mol, $g \simeq 2.02$, and $J/k_{\rm B} \simeq 37.0$~K. From the value of $C_{\rm imp}$, the impurity concentration is estimated to be $\sim 3$\%, assuming spin-$\frac12$ impurities. Below 10\,K, experimental susceptibility clearly deviates from the fit. {\cred Our microscopic analysis suggests that interchain interactions do not exceed 5\,K (Sec.~\ref{sec:dft}). Therefore, the deviations are most likely caused by anisotropic terms in the spin Hamiltonian, such as Dzyloshinskii-Moriya interaction, which is allowed by symmetry for the intrachain coupling $J_3$.}

\begin{figure}
	\includegraphics{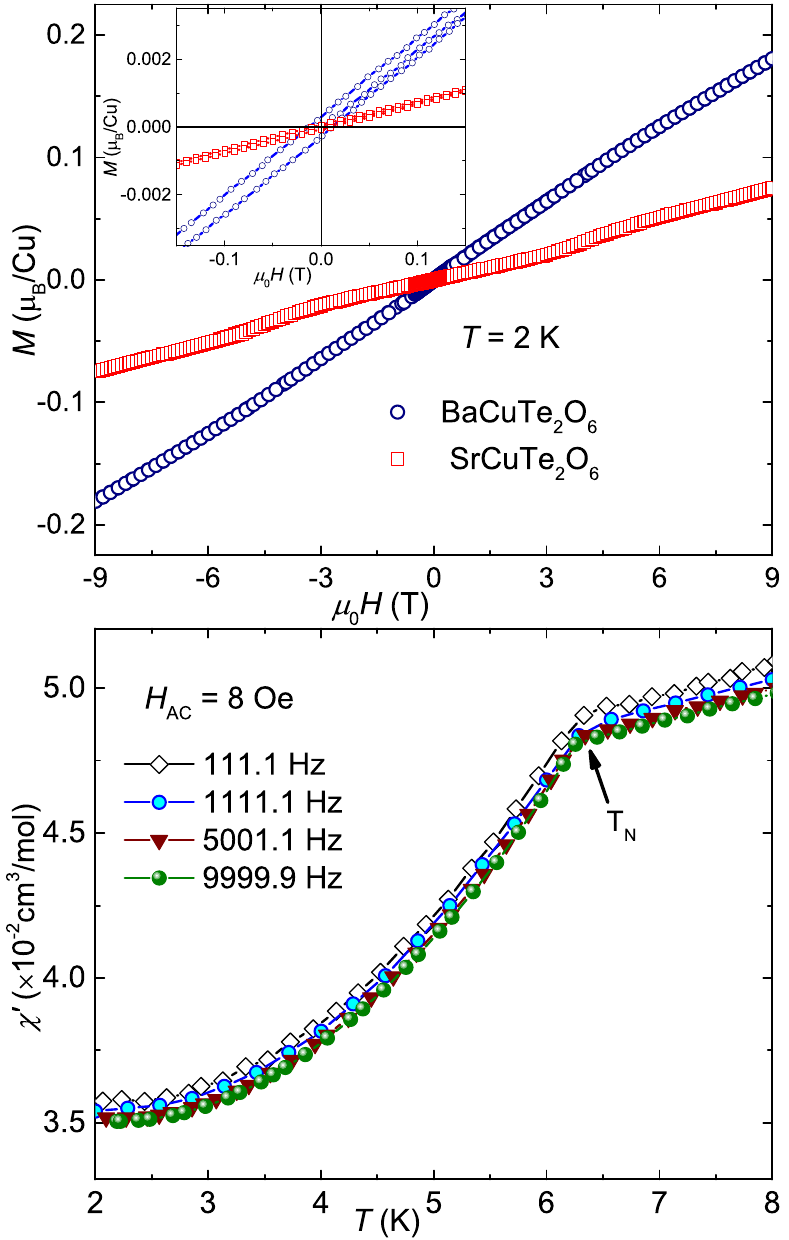}
	\caption{\label{Fig5}
		Upper panel: Magnetic isotherm ($M$ vs $H$) for BaCuTe$_2$O$_6$ and SrCuTe$_2$O$_6$ at $T = 2$~K. The bend around 3\,T in the Sr compound indicates the transition toward a field-induced state~\cite{Chillal2020b}. Inset: Magnified $M$ vs $H$ near zero field to highlight the minute net moment present in the Ba compound. Lower panel: Real part of the AC susceptibility ($\chi^\prime$) vs $T$ measured on BaCuTe$_2$O$_6$ at different frequencies. The transition temperature is marked by an arrow.}
\end{figure}

The Sr compound showed two consecutive transitions as a function of temperature in the magnetic susceptibility and other thermodynamic probes~\cite{Ahmed2015,Koteswararao2015,Chillal2020b}. In contrast, BaCuTe$_2$O$_6$ reveals one transition only (Figs.~\ref{Fig4}, \ref{Fig5}, and~\ref{Fig6}). Another difference can be seen in very low magnetic fields, where SrCuTe$_2$O$_6$ displayed no significant deviation between the field-cooled (FC) and zero-field cooled (ZFC) $\chi(T)$ measurements~\cite{Ahmed2015}, while BaCuTe$_2$O$_6$ displays a bifurcation at $T_{\rm N}$ suggestive of a weak spin canting in the magnetically ordered state (see the inset of the upper panel of Fig.~\ref{Fig4}). 

The magnetization isotherm ($M$ vs $H$) at $T = 2$~K measured upto 9~T is shown in the upper panel of Fig.~\ref{Fig5} for both BaCuTe$_2$O$_6$ and SrCuTe$_2$O$_6$. Unlike the Sr compound, the Ba compound exhibits linear behavior without any indications of a field-induced transition. In the inset of Fig.~\ref{Fig5}, we magnified the data near zero field in order to highlight a tiny remanent magnetization of less than 0.001\,$\mu_B$/f.u. present in the Ba compound. This indicates a small spin canting in the antiferromagnetically ordered state. 

AC susceptibility as a function of temperature was also measured at different frequencies. The real part of the AC susceptibility ($\chi^\prime$) shows a peak at $T_{\rm N}$, its position not changing with frequency (lower panel of Fig.~\ref{Fig5}). This excludes freezing effects. We also note that the transition temperature is remarkably insensitive to the applied field (lower panel of Fig.~\ref{Fig4}). 

\subsection{Heat Capacity}
\begin{figure}
\includegraphics{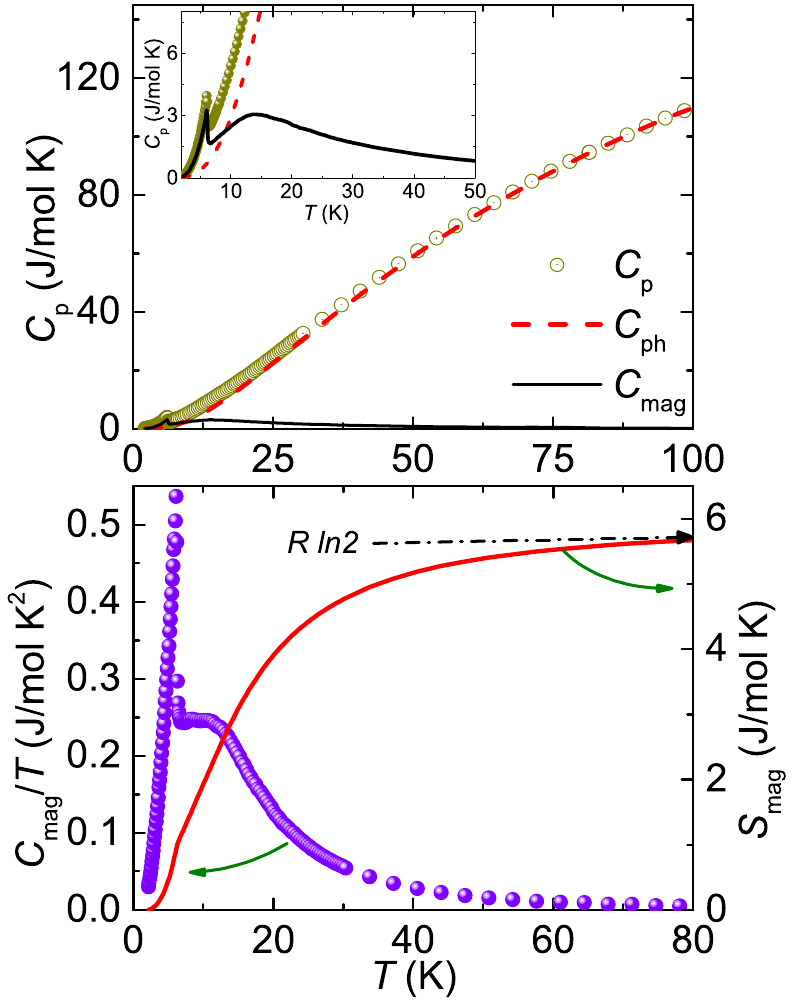}
\caption{\label{Fig6} Upper panel: $C_{\rm p}$ vs $T$ of BaCuTe$_{2}$O$_{6}$ measured in zero applied field. The dashed and solid lines represent the phononic ($C_{\rm ph}$) fit [Eq.~(\ref{Debye})] and magnetic ($C_{\rm mag}$) parts of $C_{\rm p}(T)$, respectively. In the inset, $C_{\rm mag}(T)$ is shown in the magnified form. Lower panel: $C_{\rm mag}/T$ and $S_{\rm mag}$ as a function of $T$ are shown in the left and right $y$-axes, respectively. The dash-dotted arrow points to the expected value of magnetic entropy $S_{\rm mag} = R\ln2$ for Cu$^{2+}$ ($S=\frac{1}{2}$).}
\end{figure}
Heat capacity $C_{\rm p}$ as a function of temperature measured in zero field is shown in Fig.~\ref{Fig6}. With decreasing temperature, it decreases rapidly and then shows a sharp $\lambda$-type anomaly at $T_{\rm N} \simeq 6.1$~K, affirming the transition to the magnetically ordered state. In magnetic insulators, $C_{\rm p}$ has two principal contributions, phononic part and magnetic part that dominate the signal at higher and lower temperatures, respectively.

To estimate the phononic part of the heat capacity ($C_{\rm ph}$), the raw data at high temperatures were fitted by a linear combination of four Debye functions~\cite{Gopal1966,Kittel2005},
\begin{equation}
\label{Debye}
C_{\rm ph}(T) = 9R\times\sum\limits_{\rm n=1}^{4} c_{\rm n} \left(\frac{T}{\theta_{\rm Dn}}\right)^3 \int_0^{\frac{\theta_{\rm Dn}}{T}} \frac{x^4e^x}{(e^x-1)^2} dx.
\end{equation}
Here, $R=8.314$~J\,mol$^{-1}$\,K$^{-1}$ is the universal gas constant. The coefficients $c_{\rm n}$ represent the groups of distinct atoms in the crystal, and $\theta_{\rm Dn}$ are the corresponding Debye temperatures. Since there are four different atoms with largely different atomic mass, we used four ($n=4$) Debye functions. The solid dashed curve in Fig.~\ref{Fig6} (upper panel) represents the Debye fit ($C_{\rm ph}$) of the $C_{\rm p}(T)$ data at high temperatures ($T \geq 50$~K) using Eq.~\eqref{Debye} and is extrapolated down to low temperatures. 

\begin{figure}
\includegraphics{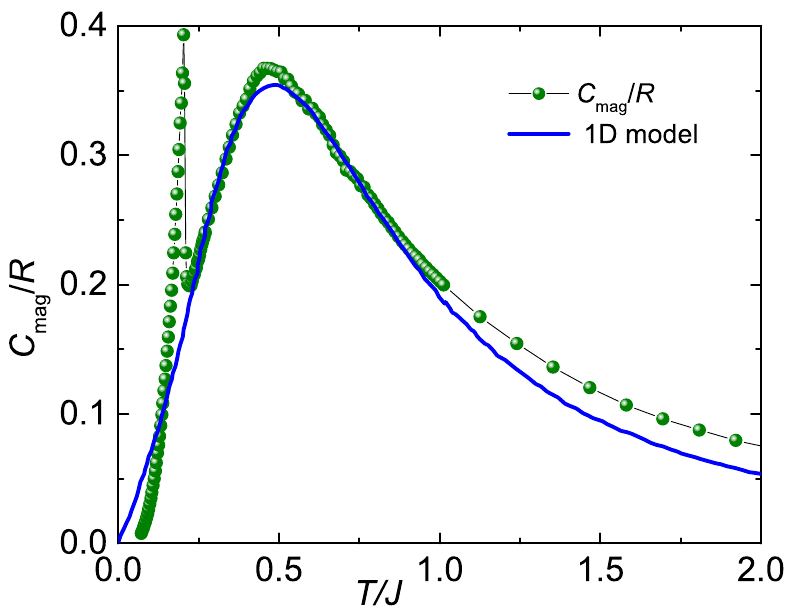}
\caption{\label{Fig7} $C_{\rm mag}/R$ vs $T$ of BaCuTe$_{2}$O$_{6}$ in zero magnetic field. The solid line is the series expansion result for the 1D spin-$\frac12$ chain model from Ref.~\cite{Johnston2000} with the exchange coupling $J/k_{\rm B} \simeq 30$~K.}
\end{figure}

Magnetic part of the heat capacity ($C_{\rm mag}$) was obtained by subtracting $C_{\rm ph}$ from the total heat capacity $C_{\rm p}$. The subtraction procedure was verified by calculating the magnetic entropy $S_{\rm mag}$ through the integration of $C_{\rm mag}(T)/T$ that yields $S_{\rm mag} \simeq 5.87$~J\,mol$^{-1}$\,K$^{-1}$ at 80~K (lower panel of Fig.~\ref{Fig6}). This value is close to the expected magnetic entropy for spin-$\frac12$: $S_{\rm mag} = R\ln 2 = 5.76$~J~mol$^{-1}$~K$^{-1}$.
In order to highlight the low-temperature features, we magnified the low-temperature data in the inset of Fig.~\ref{Fig6}. $C_{\rm mag}$ shows a clear broad maximum at $T_{\rm C}^{\rm max} \simeq 13.6$~K representing the short-range AFM correlations similar to that observed in $\chi(T)$. Both shape and magnitude of the maximum compare favorably with theoretical results for the spin-chain model (Fig.~\ref{Fig7})~\cite{Johnston2000}.

\begin{figure}
	\includegraphics{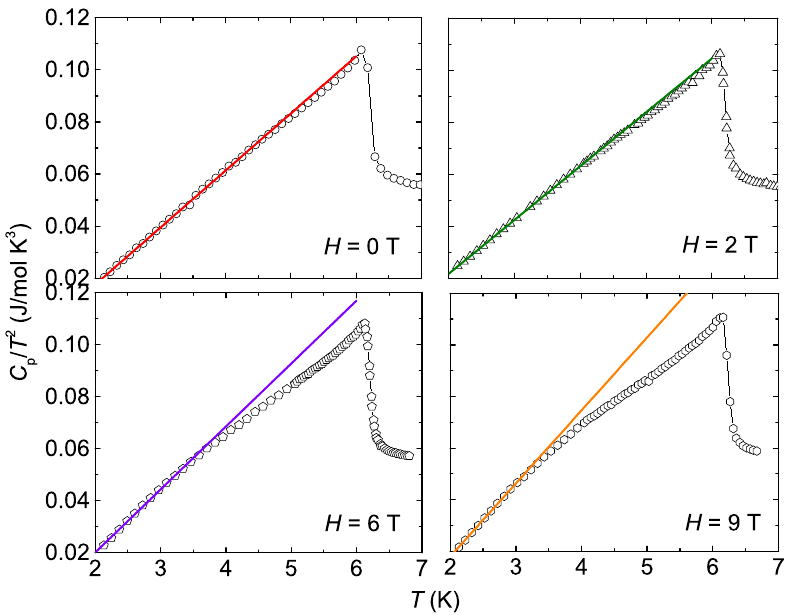}
	\caption{\label{Fig8} $C_{\rm p}/T^{2}$ vs $T$ measured under different magnetic fields from 0 to 9~T. Solid line represents the fit by $C_{\rm p} \propto T^{3}$ in the low-temperature region, well below $T_{\rm N}$.}
\end{figure}

Figure~\ref{Fig8} shows $C_{\rm p}/T^{2}$ vs $T$ measured in different applied fields at low temperatures. The value of $T_{\rm N}$ is found to remain unchanged under external magnetic field up to 9~T similar to that observed in $\chi(T)$. Below $T_{\rm N}$, $C_{\rm p}(T)$ follows a $T^{3}$ behaviour at all fields, as expected for a 3D Heisenberg antiferromagnet. This indicates the smallness (or absence) of an anisotropy gap in the spin-wave spectrum.

\subsection{Microscopic magnetic model}
\label{sec:dft}
We now compare and contrast BaCuTe$_2$O$_6$ with its Sr analog. To this end, magnetic interactions are calculated for the experimental crystal structures of both compounds determined under exactly the same conditions, by refining synchrotron XRD data collected at 80\,K (the structural parameters for SrCuTe$_2$O$_6$ are given in the Appendix). This allows us to trace relevant trends even for weak interchain couplings~\footnote{Note that the values reported in this manuscript for SrCuTe$_2$O$_6$ may be slightly different from those in Ref.~\cite{Ahmed2015}, because different structural parameters have been used.}.

\begin{table}
\caption{\label{tab:exchange}
Interatomic distances $d_i$ (in\,\r A), hopping parameters $t_i$ (in\,meV), and exchange integrals $J_i$ (in\,K) for SrCuTe$_2$O$_6$ (upper rows) and BaCuTe$_2$O$_6$ (lower rows). The interatomic distances are given according to the experimental crystal structures of both compounds determined at 80\,K, only the shortest O$\cdots$O distance is listed. AFM contributions to the exchange couplings are obtained as $J_i^{\rm AFM}=4t_i^2/U_{\rm eff}$ with $U_{\rm eff}=4.5$\,eV. Total exchange integrals $J_i$ are obtained by a GGA+$U$ mapping procedure. The interactions beyond $J_6$ are negligible.}
\begin{ruledtabular}
\begin{tabular}{c@{\hspace{2em}}cc@{\hspace{2em}}rc@{\hspace{2em}}r@{\hspace{1em}}c}
 & $d_{\rm Cu-Cu}$ & $d_{\rm O\cdots O}$ & $t_i$ & $J_i^{\rm AFM}$ & $J_i$ \\\hline
 $J_1$ & 4.530 & 3.262 & 10    & 1  & 0 & Sr \\
       & 4.736 & 3.412 & 10    & 1  & 0 & Ba \\\hline
 $J_2$ & 5.511 & 2.747 & 25    & 7  & $-5$ & Sr \\
       & 5.656 & 2.772 & 46    & 22 & 5 & Ba \\\hline
 $J_3$ & 6.276 & 2.773 & $-87$ & 78 & 47 & Sr \\
       & 6.482 & 2.927 & $-73$ & 55 & 34 & Ba \\\hline
 $J_4$ & 7.453 & --    & 0     & 0  & 0 & Sr \\
       & 7.704 & --    & $-2$  & 0  & 0 & Ba \\\hline
 $J_5$ & 8.740 & --    & 0     & 0  & 0 & Sr \\
       & 9.034 & --    & 0     & 0  & 0 & Ba \\\hline
 $J_6$ & 8.949 & --    & $-20$ & 4  & 2 & Sr \\
       & 9.196 & --    & $-18$ & 4  & 2 & Ba \\
\end{tabular}
\end{ruledtabular}
\end{table}

\begin{figure}
\includegraphics{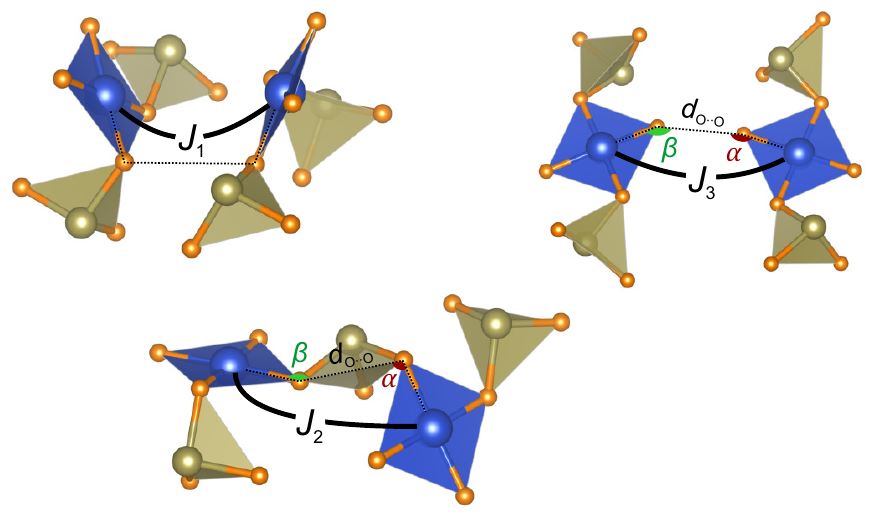}
\caption{\label{Fig9}
Superexchange pathways for the interactions $J_1-J_3$. The Cu--O--O angles $\alpha$ and $\beta$ control the relative strengths of $J_2$ and $J_3$. The O$\cdots$O distances are given in Table~\ref{tab:exchange}.}
\end{figure}

Exchange parameters $J_{ij}$ of Eq.~\eqref{eq:ham} are obtained via two complementary procedures. First, the uncorrelated GGA band structure is mapped onto a tight-binding model and supplied with an effective Coulomb repulsion $U_{\rm eff}=4.5$\,eV~\cite{Janson2011,Lebernegg2013} that yields AFM exchange terms as $J_i^{\rm AFM}=4t_i^2/U_{\rm eff}$, where $t_i$ are the hopping parameters. This way, all relevant exchange couplings, including those between distant Cu atoms, can be assessed simultaneously. Alternatively, total exchange couplings, including also ferromagnetic terms, were extracted from GGA+$U$ total energies by a mapping procedure~\cite{Xiang2011,Tsirlin2014}. 

The results of both methods are listed in Table~\ref{tab:exchange}. They suggest that both compounds are dominated by the third-neighbor interactions $J_3$ responsible for the formation of spin chains. The magnitude of $J_3$ decreases upon replacing Sr with Ba. Microscopically, the coupling $J_3$ originates from the \mbox{Cu--O$\cdots$O--Cu} superexchange controlled by the O$\cdots$O distance (Fig.~\ref{Fig9}). Such couplings usually occur through the O--O edge of a rigid polyanionic group, which is rather insensitive to pressure effects or chemical substitutions~\cite{Lebernegg2011}. However, in $A$CuTe$_2$O$_6$ the O$\cdots$O contact of $J_3$ does not form the edge of the TeO$_3$ pyramid, so the relevant distance can be quite flexible and increases by about 0.15\,\r A upon replacing Sr with Ba, thus reducing $J_3$ (Table~\ref{tab:exchange}). 

The interchain couplings $J_1$ and $J_2$ follow a similar superexchange mechanism, but with a largely different interaction geometry (Fig.~\ref{Fig9}). While $J_1$ is negligible in both compounds, we find a notable increase in $J_2^{\rm AFM}$ and, correspondingly, a change of sign of $J_2$ from FM to AFM upon replacing Sr with Ba. Here, the O$\cdots$O distance is constrained by the edge of the TeO$_3$ pyramid and changes only weakly, but the linearity of the pathway gauged by the \mbox{Cu--O--O} angles $\alpha$ and $\beta$ increases: compare $\alpha_{\rm Sr}=100.6^{\circ}$ and $\beta_{\rm Sr}=155.2^{\circ}$ to $\alpha_{\rm Ba}=106.1^{\circ}$ and $\beta_{\rm Ba}=159.9^{\circ}$. In the case of $J_3$, $\alpha_{\rm Sr}=\beta_{\rm Sr}=154.5^{\circ}$ and $\alpha_{\rm Ba}=\beta_{\rm Ba}=153.7^{\circ}$ give rise to much more ``straight'' superexchange pathways and cause a much stronger AFM superexchange than $J_2$, despite the longer O$\cdots$O separation. This structural difference elucidates the $J_3\gg |J_2|$ regime observed in both Sr and Ba compounds. 

Similar arguments can be invoked to explain the absence of $J_1$, where the O$\cdots$O distance exceeds 3.2\,\r A (Table~\ref{tab:exchange}), while both $\alpha$ and $\beta$ are in the range of $100-110^{\circ}$ and lead to a strong deviation of the pathway from linearity. Such interaction geometry is prohibitive for the superexchange (Fig.~\ref{Fig9}), similar, for example, to fedotovite and other Cu$^{2+}$ minerals with hexamer clusters~\cite{Nekrasova2020}. On the other hand, the long-range interaction $J_6$ appears to be non-negligible and comparable in size in both Sr and Ba compounds (Table~\ref{tab:exchange}).

\begin{figure}
\includegraphics{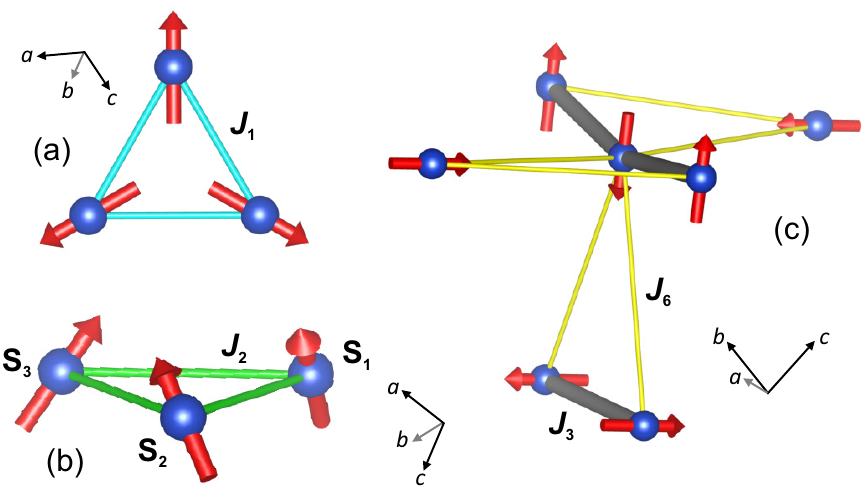}
\caption{\label{Fig10}
Frustrated loops formed by the interchain interactions $J_1$, $J_2$, and $J_6$. Experimental magnetic structure of SrCuTe$_2$O$_6$~\cite{Chillal2020b} is superimposed. {\cred We expect that BaCuTe$_2$O$_6$ has a different spin arrangement according to the different sign of $J_2$ predicted in our work.}}
\end{figure}

We infer that SrCuTe$_2$O$_6$ and BaCuTe$_2$O$_6$ should host similar spin lattices where $J_1$ is nearly absent, whereas $J_2$ changes sign from FM for $A={\rm Sr}$ to AFM for $A={\rm Ba}$. This scenario sheds new light on the magnetic structure that has been determined for SrCuTe$_2$O$_6$ recently. The neutron diffraction studies of Refs.~\cite{Chillal2020b,Saeaun2020} revealed that the spins are antiparallel along the chains, but adopt different directions in different chains. Altogether, six sublattices with different spin directions are formed. The spin direction changes not only between every two orthogonal chains, but also between the two adjacent chains that are parallel to each other [Fig.~\ref{Fig10}(c)]. 

In SrCuTe$_2$O$_6$, this strongly non-collinear spin arrangement gives rise to the coplanar $120^{\circ}$ magnetic order on the triangles formed by $J_1$, and the non-coplanar order on the triangles formed by $J_2$ [Fig.~\ref{Fig10}(a),(b)]. At first glance, it would indicate that AFM $J_1$ should be the leading interchain coupling, whereas $J_2$ is negligible, because any sizable $J_2$ would cause a different type of order on the $J_2$-triangles: collinear ferromagnetic at $J_2<0$ and coplanar $120^{\circ}$ at $J_2>0$. Interestingly, we find a sizable $J_2$ and neither microscopic indications nor plausible structural reasons for $J_1$ being the leading interchain coupling in any of the compounds. On the other hand, $J_6$ is a plausible reason for non-collinearity. This coupling connects a spin of one chain to two adjacent spins in the parallel chain (Fig.~\ref{Fig10}c). The resulting frustration is alleviated by arranging the spins of these chains along two orthogonal directions, similar to the non-collinear magnetic order in Cu$_2$GeO$_4$~\cite{Badrtdinov2019}. 

The frustrated nature of $J_6$ prevents the spins on the $J_2$ triangles from being parallel, as FM $J_2$ would require. However, partial FM correlations still occur, with the classical interaction energy,
\begin{equation}
 E_2=J_2(\mathbf S_1\mathbf S_2+\mathbf S_2\mathbf S_3+\mathbf S_1\mathbf S_3)=\tfrac32 J_2S^2
\end{equation} 
that allows a sizable energy gain from FM $J_2$. This energy gain is in fact equal to the energy gain expected from the $120^{\circ}$ order on the $J_1$ triangle, should the interaction $J_1$ be AFM in nature. Therefore, AFM $J_1$ and FM $J_2$ are equally instrumental in stabilizing the non-collinear magnetic order in SrCuTe$_2$O$_6$. 

A somewhat similar magnetic structure can be envisaged in BaCuTe$_2$O$_6$. Indeed, changing the spin directions on a given $J_2$ triangle from
\begin{equation*}
 \mathbf S_1\,\|\, [1\,\bar 1\,0],\quad \mathbf S_2\,\|\, [1\,0\,\bar 1],\quad \mathbf S_3\,\|\,[0\,\bar 1\,\bar 1]
\end{equation*}
in SrCuTe$_2$O$_6$ [Fig.~\ref{Fig8}(b)] to 
\begin{equation*}
 \mathbf S_1\,\|\, [1\,\bar 1\,0],\quad \mathbf S_2\,\|\, [\bar 1\,0\,1],\quad \mathbf S_3\,\|\,[0\,1\,\bar 1]
\end{equation*}
would result in $E_2=-\frac32J_2S^2$ and allow the energy gain from AFM $J_2$. {\cred This will lead to a different magnetic symmetry.} It would be interesting to test this prediction experimentally if neutron data of the quality sufficient to discriminate between different magnetic symmetries could be obtained for BaCuTe$_2$O$_6$. 

\subsection{Magnetic dilution}
\begin{figure}
	\includegraphics{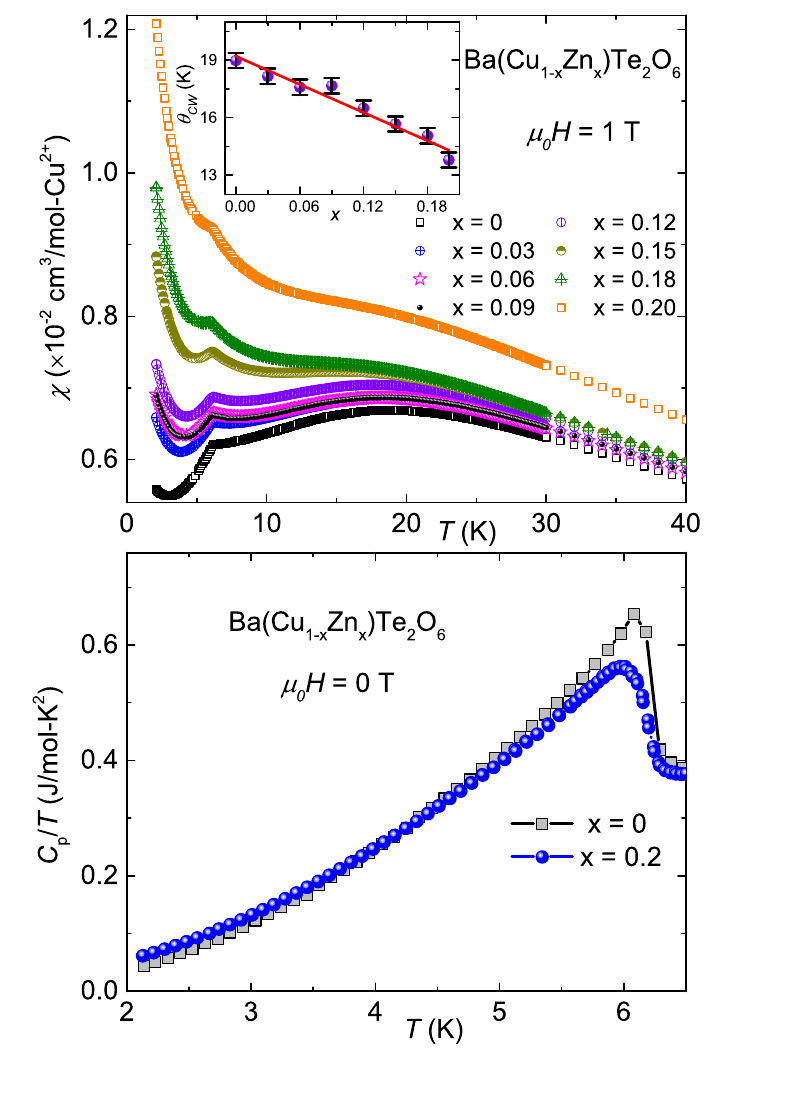}
	\caption{\label{Fig11} Upper panel: $\chi(T)$ of Ba(Cu$_{1-x}$Zn$_{x}$)Te$_{2}$O$_{6}$ samples measured at $H = 1$~T. Inset: Curie-Weiss temperature $\theta_{\rm CW}$ as a function of doping concentration $x$ with the solid line representing the linear fit. Lower panel: $C_{\rm p}/T$ of two end samples ($x = 0$ and 0.2) measured in zero magnetic field.}
\end{figure}
{\cred Zn substitution can have two-fold effect on the magnetism of BaCuTe$_2$O$_6$. The change in the cubic lattice parameter tunes chemical pressure, whereas the introduction of non-magnetic Zn$^{2+}$ ions dilutes the spin lattice. Figure~\ref{Fig3} shows that changes in the lattice parameter are minimal, so the dilution effect should prevail.} In the upper panel of Fig.~\ref{Fig11}, $\chi(T)$ of Zn-doped samples is plotted after normalizing to one mole of Cu$^{2+}$ spins. Clearly, $\chi(T)$ at low temperatures increases systematically with increasing $x$. In spin systems, when non-magnetic Zn$^{2+}$ is doped at the magnetic Cu$^{2+}$ site, free Cu$^{2+}$ ions are produced and cause the low-temperature paramagnetic upturn in $\chi(T)$. Additionally, magnetic dilution may create finite segments of spin chains that will also give rise to the Curie-like behavior~\cite{Eggert1995}. Thus, our experimental observations are consistent with the expected dilution effect.

The $\chi(T)$ data at high temperatures (above 200~K) were fitted by the CW law, Eq.~\eqref{cw}. The value of $C$ was found to vary between $\sim 0.41$ and $\sim 0.43$~cm$^{3}$K/mol-Cu$^{2+}$, which corresponds to the effective moment $\mu_{\rm eff} \simeq 1.81 - 1.85$~$\mu_{\rm B}$. The obtained value of $\theta_{\rm CW}$ is plotted as a function $x$ in the inset of Fig.~\ref{Fig11}. As the doping level increases, $\theta_{\rm CW}$ is found to decrease almost linearly from $\sim 19$~K ($x=0$) to $\sim 14$~K ($x=0.2$). This behavior is expected when some of the superexchange pathways are broken by the dilution, and the overall interaction energy gauged by $\theta_{\rm CW}$ decreases~\cite{Ranjith2015}.


{\cred N\'eel temperature should also be reduced by the dilution until long-range magnetic order disappears completely upon crossing the percolation threshold. The rate of this reduction and the value of the percolation threshold depend on the dimensionality of magnetic interactions and can be diagnostic for the interaction topology. For example, linear decrease in the $T_N$ is expected for quasi-2D antiferromagnets~\cite{Chernyshev2001}, whereas superlinear decrease indicates further reduction in the dimensionality toward 1D~\cite{Eggert2002} or an additional magnetic frustration in 2D~\cite{Carretta2011}. Interestingly, BaCuTe$_2$O$_6$ follows neither of these scenarios. Its} N\'eel temperature is remarkably insensitive to the dilution. Both the kink in $\chi(T)$ and the $\lambda$-type anomaly in $C_{\rm p}(T)$ remain around 6.1\,K even at $x=0.2$ (Fig.~\ref{Fig11}). {\cred To our knowledge, it is the only example of a Cu$^{2+}$-based system, where long-range magnetic order is essentially unaffected by the Zn$^{2+}$ substitution. One may understand this effect as a consequence of magnetic frustration for the interachain couplings, as discussed in Sec.~\ref{sec:dft}.} Dilution may alleviate the frustration of interchain couplings and keep $T_{\rm N}$ around the same value even though the overall energy of magnetic interactions is decreased.

\section{Conclusions}
We have studied the effect of negative pressure on the frustrated network of the $A$CuTe$_2$O$_6$ compounds. Despite a significant increase in the cubic lattice parameter and individual Cu--Cu distances, the newly prepared BaCuTe$_2$O$_6$ strongly resembles its Sr sibling and shows a similar quasi-1D magnetic behavior driven by the dominant third-neighbor couplings $J_3$ forming spin chains. A closer examination of individual magnetic couplings revealed that the stretching of superexchange pathways caused by negative pressure not only reduces $J_3$, but also increases AFM contribution to the second-neighbor coupling $J_2$ and likely changes its sign. This modification of the interchain coupling may be the cause of the subtle differences between the Sr and Ba compounds: the latter lacks the second phase transition at $T_{\rm N2}$ as well as the field-induced transition around 3\,T, but reveals minute spin canting. 

We have also shown that the combination of frustrated interchain couplings $J_2$ and $J_6$ may stabilize the non-collinear magnetic structure observed experimentally in SrCuTe$_2$O$_6$~\cite{Chillal2020b,Saeaun2020}. This scenario allows a generalized description of the Sr and Ba compounds in terms of a $J_2-J_3-J_6$ magnetic model, with the nearest-neighbor coupling $J_1$ systematically absent in both compounds due to its highly unfavorable superexchange geometry. The 3D spin-liquid behavior of PbCuTe$_2$O$_6$ requires the increase in AFM $J_2$ with the simultaneous reduction in $J_3$ and re-activation of $J_1$~\cite{Chillal2020a}. The structural origin of these rather striking changes would be an interesting direction for future studies.

{\cred \textit{Note added:} After completing this work, we became aware of an independent study of BaCuTe$_2$O$_6$~\cite{Samartzis2020,Samartzis2021}. Neutron diffraction experiments confirm different magnetic symmetries of SrCuTe$_2$O$_6$ and BaCuTe$_2$O$_6$, in agreement with our conclusion on the different sign of $J_2$ in these compounds.}

\acknowledgments
PB, NA, VS, and RN acknowledge SERB, India for financial support bearing sanction order no. CRG/2019/000960. The work in Augsburg was supported by the German Research Foundation (DFG) via the Project No. 107745057 (TRR80).

\appendix
\section{Crystallographic data for SrCuTe$_2$O$_6$}
Here, in Table~\ref{tab:sr}, we present structural parameters for SrCuTe$_2$O$_6$ determined at 80\,K using the same experimental setup as in the case of BaCuTe$_2$O$_6$ (Sec.~\ref{sec:methods}). The powder sample of the Sr compound from Ref.~\cite{Ahmed2015} was used for the synchrotron XRD measurement. The results are only marginally different from the previous reports~\cite{Wulff1997,Chillal2020b,Saeaun2020}, and facilitate a direct comparison between the two compounds.

\begin{table}
\caption{\label{tab:sr}
Atomic coordinates and atomic displacement parameters ($U_{\rm iso}$, in units of $10^{-2}$\,\r A$^2$) for SrCuTe$_2$O$_6$ at 80\,K. The lattice parameter is $a=12.4406(1)$\,\r A, and the space group is $P4_132$. 
}
\begin{ruledtabular}
\begin{tabular}{cccccc}
 Atom & Site & $x/a$ & $y/b$ & $z/c$ & $U_{\rm iso}$ \\\hline
 Sr1 & $8c$ & 0.05417(4) & 0.05417(4) & 0.05417(4) & 0.43(2) \\
 Sr2 & $4b$ & 0.375      & 0.625      & 0.125      & 0.59(3) \\
 Cu  & $12d$ & 0.47634(7) & 0.875      & 0.27366(7) & 0.48(3) \\
 Te  & $24e$ & 0.33846(3) & 0.91916(3) & 0.05901(3) & 0.35(1) \\
 O1  & $24e$ & 0.3722(3)  & 0.8200(3)  & 0.1713(3)  & 0.19(9) \\
 O2  & $24e$ & 0.2670(3)  & 0.8116(3)  & $-0.0227(3)$ & 0.33(9) \\
 O3  & $24e$ & 0.2202(3)  & 0.9769(3)  & 0.1302(3)  & 0.91(9) \\
\end{tabular}
\end{ruledtabular}
\end{table}

%

\end{document}